\documentclass{kapproc} 
\let\footnote\savefootnote
\let\footnotetext\savefootnotetext 
\let\footnoterule\savefootnoterule 
\kluwerbib

\RequirePackage{graphicx}%
\RequirePackage{epsf}%

\begin{document}

\articletitle{M\lowercase{g} II / C IV Kinematics vs.\ Stellar
Kinematics in Galaxies}

\author{Chris Churchill}
\affil{Penn State, University Park, PA 16803}
\email{cwc@astro.psu.edu}

\author{Chuck Steidel}
\affil{Caltech/Palomar Observatories, Pasadena, CA, 91125}
\email{ccs@astro.caltech.edu}

\begin{abstract}

Comparisons of the kinematics of {MgII} absorbing gas and the stellar
rotation curves in $0.5 \leq z \leq 1.0$ spiral galaxies suggests
that, at least in some cases, the extended gaseous envelopes are
dynamically coupled to the stellar matter.  A strong correlation
exists between the overall kinematic spread of {MgII} absorbing gas
and {CIV} absorption strength, and therefore kinematics of the
higher--ionization gas.  Taken together, the data may suggest a
``halo/disk connection'' between $z\sim 1$ galaxies and their extended
gaseous envelopes.  Though the number of galaxies in our sample are
few in number, there are no clear examples that suggest the gas is
accreting/infalling {\it isotropically\/} about the galaxies from the
intergalactic medium.

\end{abstract}

\section{Extended Gaseous Envelopes: Halos or IGM?}

For $0.5 \leq z \leq 1$, there are observed correlations between
galaxy luminous properties and {MgII} absorption properties that
support a view in which metal--enriched extended ($\sim 40$~kpc)
gaseous envelopes of normal bright galaxies are coupled to
galaxies (e.g.\ \cite{ref:bb91}; \cite{ref:sdp94};
\cite{ref:steidel95}).  An alternative view, extracted from numerical
simulations of cosmic structure growth, is that the gas is
concentrated along intergalactic filaments, where matter overdensities
also give rise to mergers and normal bright galaxies.  

By $z \sim 1$, do galaxies remain coupled to the cosmic flow of
baryons driven by matter overdensities or have they decoupled?  If the
latter, they likely sustain their gaseous envelopes via mechanical
means within the galaxies.  In this contribution, we present data that
suggest the {MgII} absorption and the emission line kinematics are
coupled in some galaxies.  We also discuss the kinematic relationship
between {CIV} and {MgII} and present the first galaxy for which data
of the emission and {MgII} and {CIV} absorption kinematics are
available.

\section{Mg II Gas--Galaxy Kinematics}

In Figure 1, we show schematics of quasar sightlines through galaxies.
Two simplified kinematic models are illustrated: isotropic infall
(left), to depict the IGM inflow, and disk rotation (right), to depict
gas coupled to the stellar components.  The sightlines pass through
absorbing clouds whose velocity vectors are shown.  Below
each schematic is an absorption profile. All profiles are in the
systemic rest--frame velocity.

\begin{figure}[th] 
\plotone{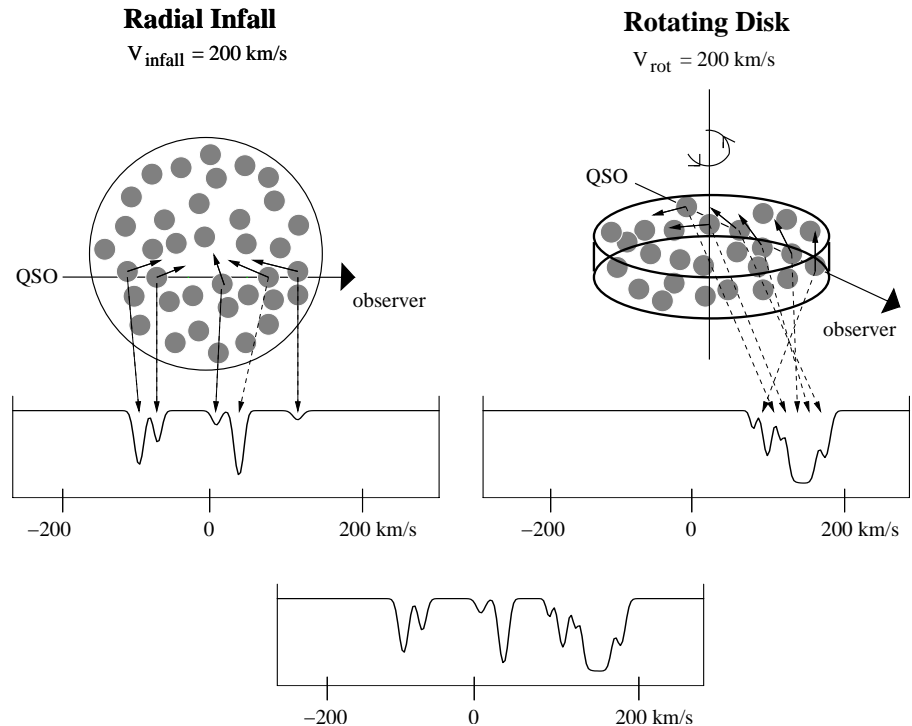} 
\caption{--- Schematic kinematic models for absorbing gas.  The left
panels show an isotropic infall model and profile (i.e.\ IGM
accretion); the right panels show the disk rotation model and profile
(i.e.\ gas coupled to stellar kinematics).  The lower panel is the
combined absorption profile.  The velocity zero point is the galaxy 
systemic redshift.}
\end{figure}

The distribution of velocities for the infall model is symmetric about
the galaxy systemic velocity and the profile is comprised of discreet,
unblended absorption lines with a velocity spread comparable to the
infall velocity.  The disk model gives rise to a profile that is
offset with both magnitude and sign dictated by rotation and is
comprised of a complex blend with a varying optical depth spread over
a narrow velocity range.  If the sightline passes through both
infalling and rotating components, the profile would appear as shown
in the bottom central panel.  These distinct absorption/kinematic
signatures of infalling and rotating gas serve as a guide for
discriminating scenarios of the nature of extended gaseous envelopes
around galaxies (for additional details see \cite{ref:kinmods}).

Steidel et~al.\ (2002) \nocite{ref:steidel02} compared the
emission--line (stellar) rotation curves of five highly inclined
spiral galaxies to their {MgII} absorption kinematics.  In four of
five galaxies, the absorption profiles are suggestive of
``disk--like'' dynamics, exhibiting properties of the disk model
(Figure 1).  The fifth galaxy exhibits a single, weak {MgII} absorber
(see \cite{ref:weakI}) at the galaxy systemic velocity.  There was no
example of discreet clouds distributed symmetrically about the galaxy
systemic redshift.  However, detailed interpretation of the absorption
kinematics is, in reality, not clear.  It is difficult to understand
the spatial geometry of the gas for the observed kinematics, the high
galaxy inclinations, and the large impact parameters.

These systems serve as evidence that, at least in some cases, the
extended gaseous envelopes around galaxies appear to be coupled to
the emission--line kinematics.  To be fair, it cannot be ruled out
that non--isotropic IGM accretion could have the general sense
of galactic rotation.

\section{Clues from C IV Absorption}

The kinematics of {MgII} and {CIV} are strongly correlated
(\cite{ref:civ-letter}).  Churchill et~al.\
suggested that this correlation could
arise if the gas spatial and kinematic distribution reflected a
disk/halo connection similar to those in local galaxies (see
\cite{ref:dahlem}, and references therein).  This is consistent with
the gas having a multiphase ionization and kinematic structure (e.g.\
\cite{ref:bergeronKP-94}; \cite{ref:archiveI},
2002b\nocite{ref:archiveII}).

To better understand this {MgII}--{CIV} kinematics correlation and the
multiphase structure, we have observed the {CIV} with STIS/HST (E230M,
$R=30,000$).  In Figure 2, we present four selected systems, S1--S4.
For each, the top panel shows the {MgII} $\lambda 2796$ transitions
(HIRES/Keck, $R=45,000$) and the lower panel shows the {CIV} $\lambda
1548$ transition in rest--frame velocity (zero points are arbitrary).
The four {MgII} profiles were specifically selected to illustrate the
disk--like kinematic signature.  Observationally, this signature is
common--- in a sample of 23 {MgII} systems, only one system exhibited
absorption symmetrically about the strong complex (\cite{ref:cv01}).

\begin{figure}[th]
\plotone{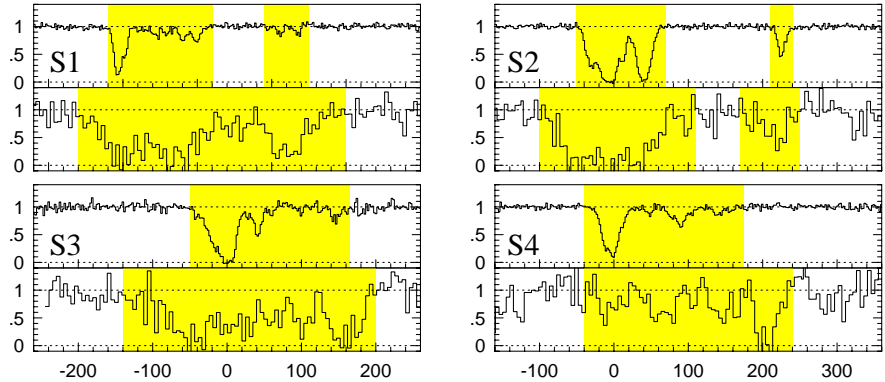}
\caption{--- The {MgII} $\lambda 2796$ (HIRES/Keck) and {CIV} $\lambda
1548$ (STIS/HST) absorption profiles for four selected systems.  The
profiles are presented in rest--frame velocity, where the zero point
has been set arbitrarily.}
\end{figure}

In systems S1 and S2, the {CIV} kinematically traces the {MgII},
though the {CIV} has less substructure (broader components).  The bulk
of the {CIV} arises in a lower density phase, possibly supported by
turbulence.  These could be coronal structures similar to that of our
Galaxy (e.g.\ \cite{ref:savage97}).  System S3 is similar to S1 and
S2, except that the strongest {CIV} component is offset in
velocity where the {MgII} is very weak.  Further, the component is
relatively narrow.  System S4 is unique in that the {CIV} is highly
structured with the {MgII} absorption but has a strong, very narrow
{CIV} component where there is no observed {MgII}.  This is a
quiescent high ionization``{CIV}--only cloud''.  The widths of these
{CIV}--only clouds would be substantially broader if they were
Galactic--like corona or shock heated infalling material.

\section{Q1317+227 at $z=0.66$: A Case Study} 

S4 is the $z=0.6610$ absorber along the Q$1317+277$ sightline.  In
Figure 3$a$, we present the WFPC2 image of the quasar field showing
two galaxies, G1 and G2.  LRIS/Keck spectra of these galaxies were
obtained with the slit aligned as shown by the vertical lines.  The
redshift of G2 matches the absorption redshift; its rotation
curve is shown in the upper panel of Figure 3$b$, below which are the
{MgII} $\lambda 2796$ and {CIV} $\lambda 1548$ absorption profiles.

\begin{figure}[bh]
\plotone{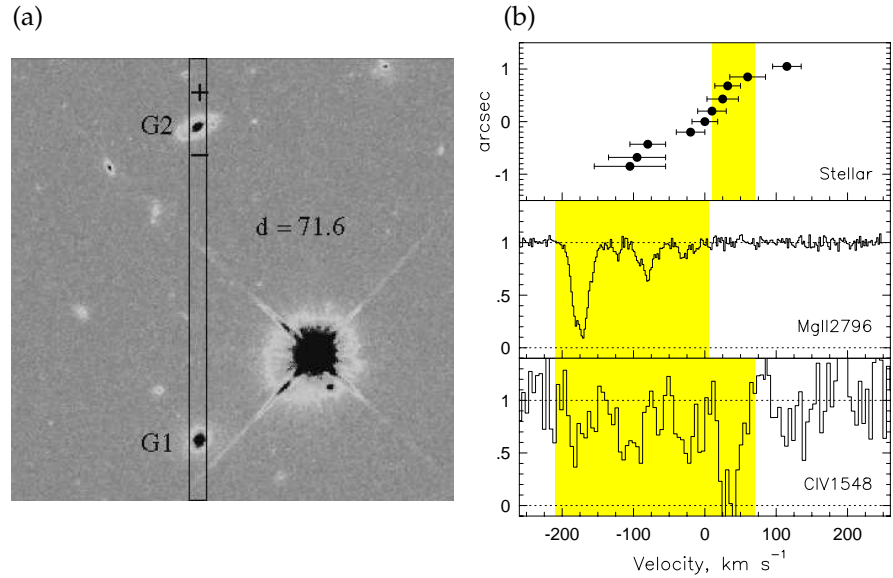}
\caption{--- ($a$) A WFP2 image of the Q$1317+277$ field, showing the
$z=0.6610$ galaxy (G2).  Vertical lines show slit
placement for the galaxy spectra.  --- ($b$, top to
bottom) The emission--line velocity profile of G2; the {MgII} $\lambda
2796$ absorption profile; and the {CIV} $\lambda 1548$ absorption
profile.  The velocity zero point is the systemic velocity of G2.}
\end{figure}

At face value, the {MgII} kinematics are suggestive of the disk
kinematic model.  The strongest {MgII} is aligned with the stellar
rotation and the weaker clouds are not symmetry about the galaxy
systemic velocity.  However, detailed modeling reveals that a simple
disk scenario fails; this system is very puzzling
(\cite{ref:steidel02}).  At a projected distance of $\sim 72$~kpc, the
nature of the narrow {CIV}--only cloud, which is slightly positive
with respect to the galaxy systemic velocity (shaded region on the
stellar velocity curve), is also
difficult to understand in view of the overall absorption kinematics.
What is the nature and origin of the quiescent high ionization gas at
a distance of 70~kpc having a nearly galactic systemic velocity?

\section{Discussion}

If galaxy evolution to the present epoch is governed by the accretion
of gas from the IGM, the gas would provide a tracer of the structure,
kinematics, and chemical enrichment of the cosmic web.  The gas would
not necessarily be coupled to galaxy emission--line kinematics in the
majority of cases; neither merging events nor IGM accretion predict
strong coupling between the gas kinematics and the stellar kinematics.
A large statistical sample is needed to discern the veracity of this
expectation.

What scenario, then, can predict the observed coupling between the
kinematics of the extended gas envelopes and the galaxy stars?
Following a merging event, star formation rates are elevated long
after the stellar system has relaxed.  Supernovae inject gas into the
halos of their host galaxies.  This scenario naturally provides for
the expulsion of gas from galaxies that is metal enriched and harbors
some memory of the dynamical state of the stellar component of the
galaxies.

\acknowledgments Supported in part by NASA NAG5 6399, NSF AST
95--96229 and AST 00--70773, {\it HST\/} GO 08672.01-A, GO
05984.01-94A, and GO 06577.01-95A, and the David \& Lucile Packard
Foundation.  We also thank K. Adelberger, J. Charlton, M. Dickinson,
B. Jannuzi, J. Maserio, M. Pettini, J. Rigby, A. Shapley,
and S. Vogt.

\end{document}